\documentclass[12pt]{iopart}

\usepackage{graphicx}

\begin{document}

\title[]{Damping and non-linearity of a levitating magnet in rotation above a superconductor}

\author{J. Druge, O. Laurent, M. A. Measson and I. Favero}

\address{Mat\'{e}riaux et Ph\'{e}nomenes Quantiques, Universit\'{e} Paris Diderot, CNRS UMR 7162, Sorbonne Paris Cit\'{e}, 10 rue Alice Domon et L\'{e}onie Duquet, 75013 Paris, France}
\ead{ivan.favero@univ-paris-diderot.fr, marie-aude.measson@univ-paris-diderot.fr}

\begin{abstract}
We study the dissipation of moving magnets in levitation above a superconductor. The rotation motion is analyzed using optical tracking techniques. It displays a remarkable regularity together with long damping time up to several hours. The magnetic contribution to the damping is investigated in detail by comparing 14 distinct magnetic configurations, and points towards amplitude-dependent dissipation mechanisms. The non-linear dynamics of the mechanical rotation motion is also revealed and described with an effective Duffing model. The obtained picture of the coupling of levitating magnets to their environment sheds light on their potential as ultra-low dissipation mechanical oscillators for high precision physics.
\end{abstract}

\maketitle

Mechanical oscillators with ultra-low dissipation find applications as frequency standards, probes of minute forces like in Atomic Force Microscopy, or signal processing where they can serve as fine radio-frequency filters. At a basic science level, they received attention in the past for their impact in high precision physics and metrology \cite{Braginsky77,Braginsky85}. More recently, they have become a central subject for a whole community of physicists aiming at observing the quantum behavior of mesoscopic mechanical systems \cite{Oconnel10,Braginsky92}. This quest has seen impressive advances notably thanks to optomechanical systems \cite{Favero09,Marquardt09,Aspelmeyer10,Marquardt12} that use the concepts of coupling light and mechanical motion, notably in the regime where the quantumness of the mechanics starts being tangible \cite{Teufel11,Chan11}. In all these situations, low dissipation of the mechanical degree of freedom is required, to protect it against quantum decoherence or classical fluctuations of its environment.

Sources of dissipation are manyfold, but one ubiquitous amongst mechanical systems is the anchoring (clamping) loss, which stems from the fact that the system is mechanically attached to a support. To circumvent this source of loss, one obvious solution is to levitate the mechanical system. If we disregard optical levitation \cite{Ashkin70,Barker10,Chang10,Romero10}, which restricts to nanoscopic mass objects, diamagnetic effects in superconductors are the most established technique to levitate a macroscopic mass and hence isolate a mechanical oscillator from its support. Superconducting magnetic levitation is been extensively studied in the context of bearings for transportations \cite{Moon94} but surprisingly, the applications of these systems in high precision physics remains relatively scarce \cite{Achilli95}. To our knowledge, there is for example no complete picture of what ultimate level of dissipation such system could reach in a refined low-amplitude force sensing experiment or in the quantum regime. This paper is a first step to try to answer these questions. To that aim, we carry out simple but systematic experiments on one of the most ubiquitous systems: a magnet levitating over a high critical temperature (Tc) superconductor.\\

\par

\begin{figure}
\includegraphics*[width=15cm]{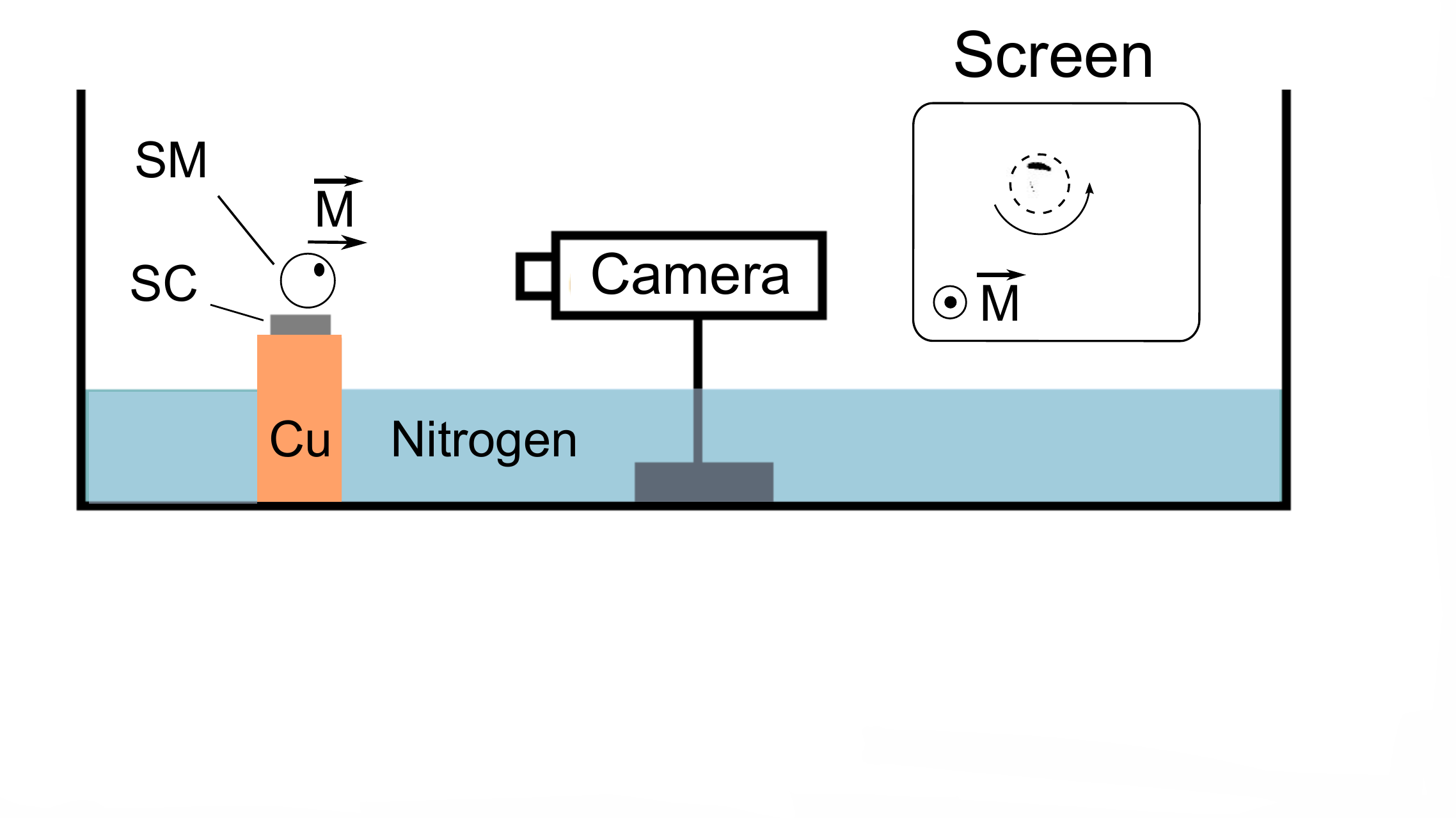}
\caption{ Schematics of the experiment. A spherical magnet (SM) is levitating and experiences a rotational motion over a superconductor (SC) pad thermally connected to a copper (Cu) cylinder plunged into liquid Nitrogen. $\vec{M}$ is the magnetic dipole moment of the magnet. Inset shows a typical video image after treatment with the mark spot (in black) within the rotating magnet boundaries (dashed circle).}
\label{fig1}
\end{figure}

We employ commercial NdFeB sphere magnets \cite{Supermagnet} of diameter varying between 5~mm and 25~mm and position them one after another over a high-Tc Y1.65Ba2Cu3O7-x superconducting cylinder pad of diameter of about 40~mm \cite{ATZ}. As illustrated in Fig.\ref{fig1}, the superconducting pad is kept at low temperature below its critical temperature Tc by contacting it with apiezon grease to a large copper cylinder immersed into a liquid nitrogen bath. In order to increase thermal inertia and stability of the set-up during hour-long measurements, a polystyrene box of very large liquid capacity is used as the bath and placed on a mechanically isolating bench. Air turbulence and convection are kept to a low level during measurements.

In this work, we focus on the rotation motion of the sphere magnet around its magnetic axis because this motion presents a very low dissipation. Indeed, in case of a perfectly homogeneous and spherical magnet, the magnetic configuration of the experiment is invariant upon rotation of the sphere around this axis. As a consequence, strictly no magnetic damping of the sphere rotation motion should occur. As we will see, even in a real and non-ideal experimental situation, this argument still holds to some extent and very long damping times of hours can be observed for this rotation motion. Note that other types of motion involving millimeter scale displacements of the magnet center of mass are systematically observed to damp rapidly in much less than a minute, and will hence not be considered here. 

A video camera is positioned near the magnet to register its motion. The camera optical axis is superposed to the sphere magnetic axis defined by the magnetic dipole $\vec{M}$. In order to allow systematic data analysis of the motion, a black mark spot is drawn on the magnet and a treatment of each video image is performed to increase its contrast. After treatment, a typical image displays the isolated black mark spot over a white background where the magnet spherical boundary is hardly visible (see inset of Fig.\ref{fig1}). This strong contrast allows for each image to run an automated search of the mark spot barycentre leading its radial coordinate (amplitude and phase) with respect to the center of the spherical boundary. In a rotation around $\vec{M}$ the amplitude remains constant and the motion is analyzed by registering the phase evolution upon time. In the following, we use this method to measure both rotation motion (with increasing phase upon time) and rotation-oscillation motion (with oscillating phase upon time). The observed motion damping is highly dependent on the orientation of the magnet dipole $\vec{M}$ with respect to the rest of the set-up with a minimum damping consistently observed when the axis is horizontal to the pad plane. We therefore adopt this orientation for all reported experiments (Fig.\ref{fig1}). Second, the damping also depends on the height at which the magnet is levitating above the superconducting pad. Hence, in our experiments we employ a constant levitation height, defined as being the distance between the upper pad plane and the bottom of the sphere. To that purpose every sphere magnet is first deposited at room temperature on an interstitial teflon element of fixed height of about 12 mm sitting on the pad. The system is then cooled below Tc and the interstitial element removed before setting the levitating magnet in rotation.

\begin{figure}
\includegraphics*[width=15cm]{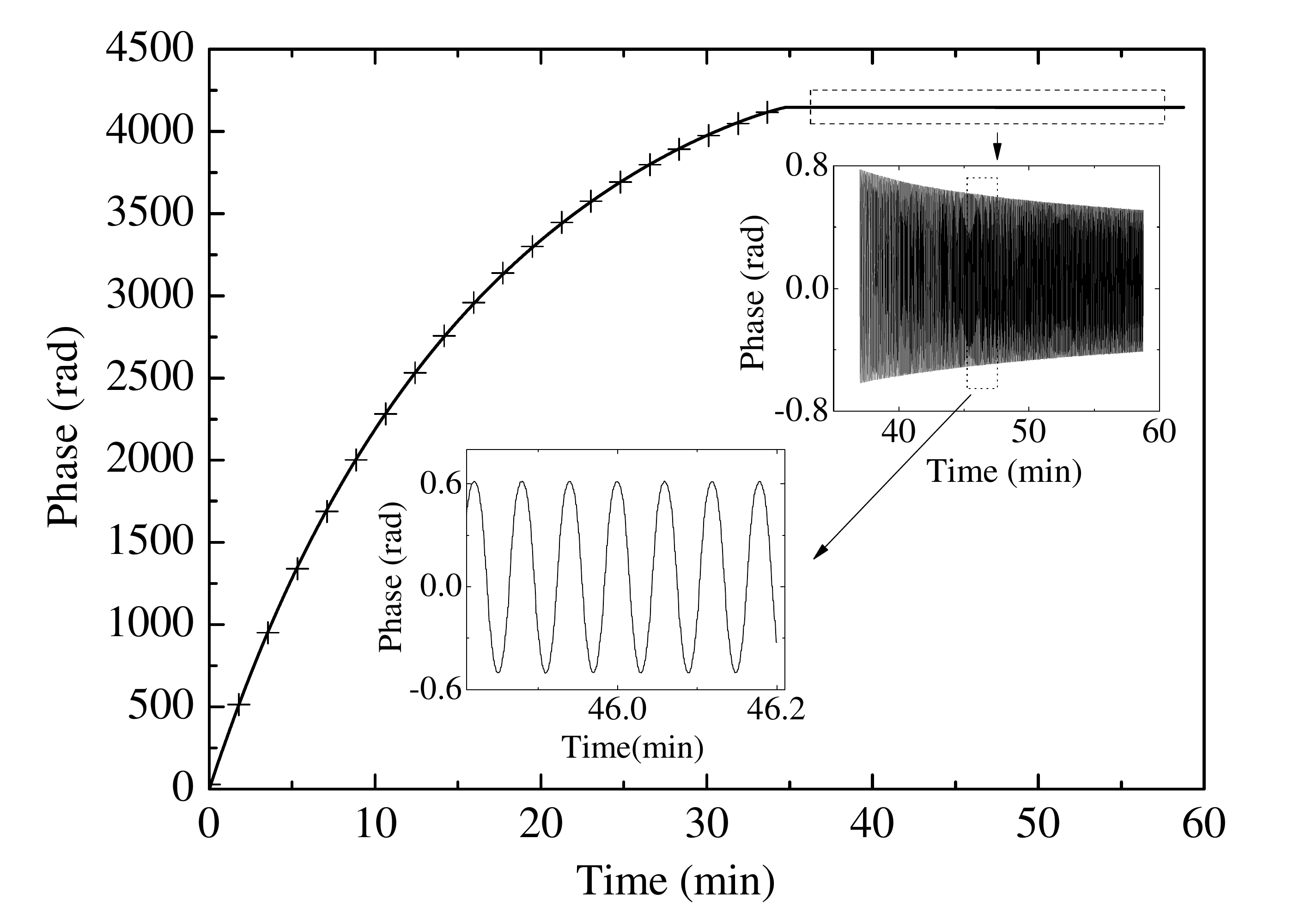}
\caption{Typical rotation motion of a levitating sphere magnet. Time evolution of the phase for a sphere magnet of diameter 12.7 mm levitating over a soft pinning superconductor. Solid lines are employed for experimental data in order to illustrate their remarkable regularity, while cross symbols are employed for the exponential fit. The top-right inset is a first close-up on the rotation-oscillation motion, where the envelope of the phase trace is apparent. The lower inset is a second close-up focusing on the time oscillation, where a time period of a few seconds is visible. }
\label{fig2}
\end{figure}

Fig.\ref{fig2} shows a typical measurement of the phase upon time after the levitating sphere magnet has been put in rotation manually. The time zero in the measurement is set by the sampling rate of the video camera ($24$~images/s), which precludes acquisition of the motion before the rotation speed has decreased down to about 48$\pi~ rad/s$. As a consequence all rotation measurements shown hereafter effectively start with the same rotation speed of about 48$\pi~ rad/s$ at time zero. In Fig.\ref{fig2}, a rotation motion first lasts about 35~min before reaching a plateau where the phase oscillates upon time corresponding to a rotation-oscillation motion, as seen clearly in the lower inset. During the first rotation part, the rotation speed progressively decreases and the phase is very accurately described by a time exponential function typical of a linear damping model. In the second part where rotation-oscillation takes place, the oscillation amplitude is itself damped in a time exponential manner as seen in the top right inset of Fig.\ref{fig2}. This behavior is again reminiscent of the harmonic oscillator model with linear damping. It is worth noting that the rotation motion of some magnets is observed to last for more than 8 hours, reaching the limit of our ability to measure it reliably. In the following we study in details the origin of the observed damping.

\begin{figure}
\includegraphics*[width=15cm]{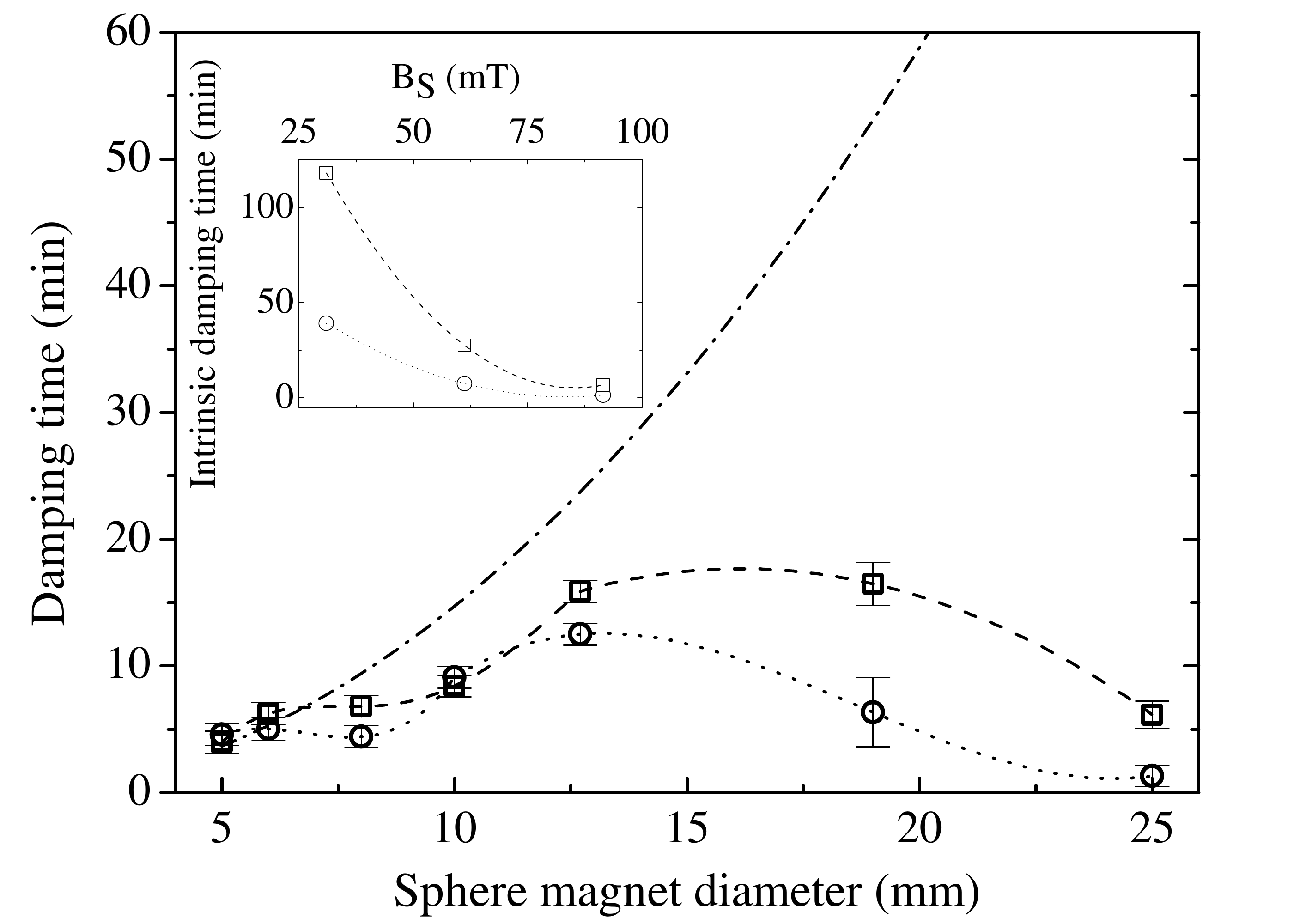}
\caption{Damping of the rotation motion for different magnetic configurations. The main plot shows the total damping time as a function of the sphere diameter for a soft pinning (squares) and hard pinning (circles) superconductor. The inset shows the intrinsic damping time as a function of the magnetic field on the top surface of the superconducting pad (see text for details), showing the role of magnetic effects in the damping.}
\label{fig3}
\end{figure}

First, as our experiments are run under ambient conditions, the surrounding air can be a source a friction for the rotation motion. However we do not expect air to produce a restoring torque, while the existence of such torque is implied by the observation of oscillatory rotation motion in the experiments. The forces responsible for this torque act at distance on the spheres and may be magnetic in nature. If true, this would contradict the picture of a rotation invariance of the sphere around its magnetic axis, which would imply no magnetic restoring torque on the rotational degree of freedom. The rotation invariance is also questioned by our independent observation of an inhomogeneous magnetic field as one rotates the magnetic sphere, which we could reveal in the orientation of iron filings and by using a magnetic foil. We conclude that the employed magnetic spheres are not perfectly rotation-invariant around their magnetic axis and that a strictly null magnetic damping of the rotation can not be expected. 

To study the role of magnetic effects in the damping of the rotation motion, we now vary both the magnetic configuration of the spheres and their magnetic environment. To that aim we vary the diameter of the spheres  and employ two types of superconductor for levitation, one with strong vortices pinning and one with soft pinning. In each configuration, we systematically measure the rotation damping during the first part of the motion where the phase decays exponentially. 

Fig.\ref{fig3} reports the measured damping time as a function of the sphere diameter, exploring 14 different configurations in total. The open square (circle) symbols correspond to the superconductor with soft (hard) pinning of the vortices. The dash-dotted line is a theoretical value for the air damping contribution, which is obtained using the Stokes model for a rotating sphere in a non-turbulent incompressible fluid \cite{BookLandau}. In Fig.\ref{fig3} the damping measured for the spheres of small diameter 5 and 6 mm seem to be explained by air damping, but as the magnet diameter increases there is a strong departure from  this contribution. Air damping has a negligible contribution on the spheres of large diameter like 12.7, 18 and 25 mm. On these spheres, the measured damping time is systematically shorter (longer) when using the strong (soft) pinning superconductor, showing that the magnetic properties of the superconductor play a key role in the rotation dynamics. To reveal even more clearly these magnetic effects, we plot in the inset of Fig.\ref{fig3} the damping time as a function of the sphere magnetic field on the superconductor pad. To that aim we measure the magnetic field of the sphere alone (with no superconductor) with a Teslameter at a distance from the sphere corresponding to the levitation height, on the magnetic equatorial line where the pad lies in the levitating configuration. The theoretical air damping contribution is removed in these data to make the "intrinsic damping time" directly appear. The plot clearly reveals that this "intrinsic damping" increases as the magnetic field on the pad increases, and as one passes from a soft pinning to a strong pinning superconductor. 

These observations ascertain the importance of vortices in the damping of rotation motion. Indeed the larger the magnetic field applied by the magnet to the superconductor pad, the larger the amount of vortices accommodated in the pad. The vortices are known to be responsible for the rigidity of the superconducting levitation configuration but this rigidity contribution is also accompanied by a dissipative contribution. In early experiments a magnet was displaced above a superconductor and lossy hysteric behavior was observed that revealed energy dissipation as vortices move in the superconductor \cite{Moon88,Moon95,Matsushita07}. In our experiments, because the sphere magnet is not perfectly symmetric, its rotation motion modulates the vortices configuration in the superconductor and produces dissipation. Our measurements show that this dissipation is stronger when there are more vortices and when the vortices are strongly pinned in the superconductor.

If the observed magnetic damping finds its roots in the vortices dissipative dynamics, it should then depend on the amplitude of the magnet motion. Indeed in case of a large motion amplitude, vortices are forced to hope between different pinning sites hence producing a strong dissipative creep. In case of a smaller amplitude, each vortex rests on its pinning site and only the dissipative part of the pinning force participates to the damping. An amplitude-dependent damping is then expected. There has been a recent strong interest in such non-linear damping mechanisms in nanoscale mechanical systems, which would make them depart from the conventional damped harmonic oscillator behavior \cite{Eichler11,Lifschitz08}. Here we will not specifically focus on these non-linear aspects of the damping but show on a more general foot that strong non-linearities are indeed present in the dynamical behavior of rotating magnets in levitation above superconductors.

\begin{figure}
\includegraphics*[width=15cm]{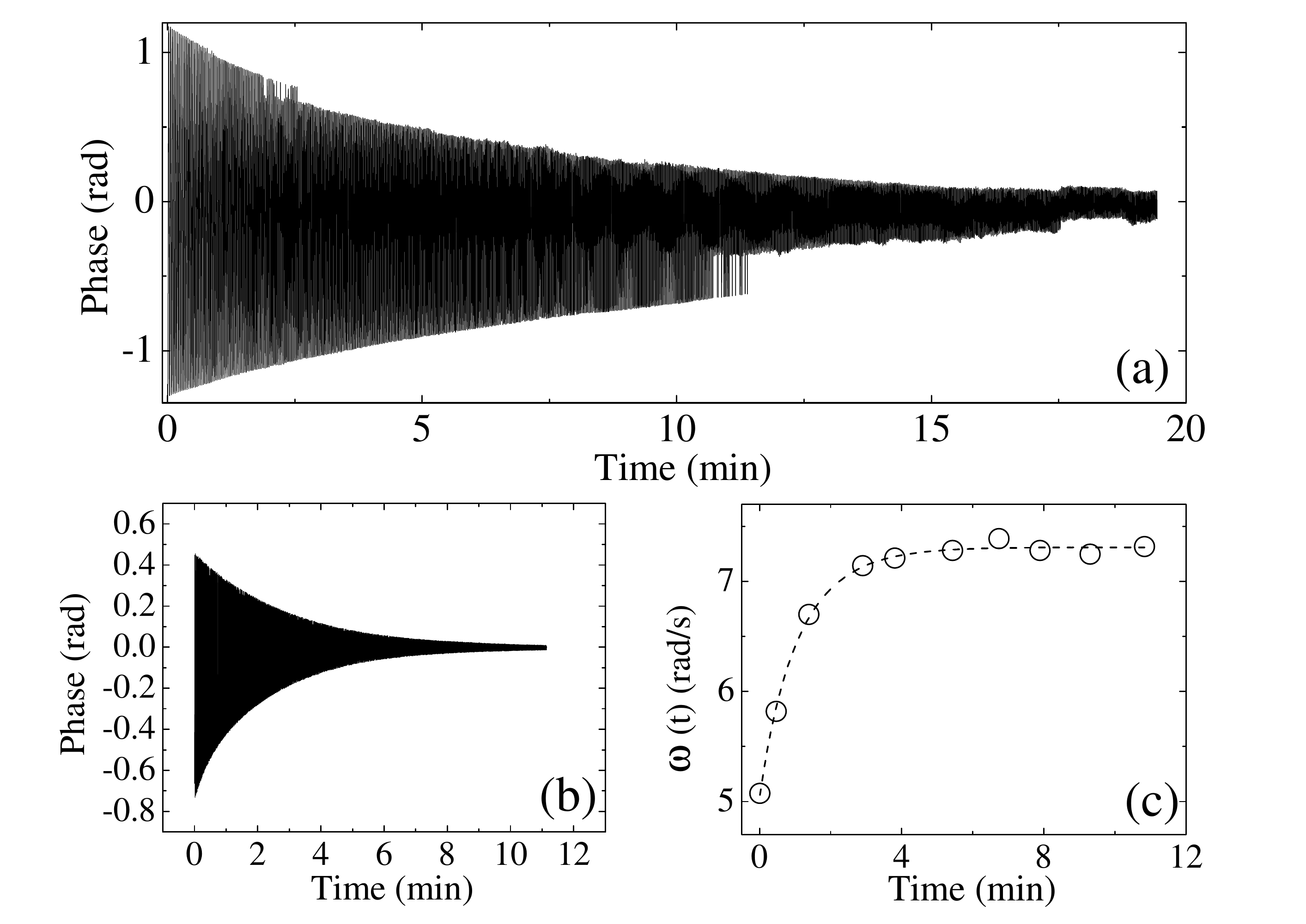}
\caption{Non linearities in the rotation of magnets levitating above a superconductor. (a) shows the time evolution of the phase of a rotation-oscillation motion, in a case where the envelope experiences several abrupt changes upon time. (b) is the final evolution after the last abrupt change, displaying an asymmetry of the envelope. (c) is the corresponding time evolution of the instantaneous angular frequency. The open circles are data and the dashed line is the fit function predicted by the effective Duffing model (see text for details).}
\label{fig4}
\end{figure}

Fig.\ref{fig4}a shows the phase time evolution of a 19 mm diameter sphere magnet in rotation-oscillation above a strong pinning superconductor. The evolution is qualitatively different from the one shown in Fig.\ref{fig2}, in that several abrupt changes are now visible in the envelope evolution. These abrupt changes cannot be explained by an harmonic oscillator model and convey the picture of an oscillation motion within multiple adjacent potential wells. As the mechanical energy dissipates upon time, the system progressively restricts its motion to fewer wells until it resides within a single of these, where the mechanical energy finishes to be dissipated. Fig.\ref{fig4}b shows such final evolution in the last well for a 25 mm diameter sphere levitating over the same superconductor. Even in this case where a unique well is involved, the phase evolution reveals a non-linearity. The envelope amplitude is strongly asymmetric with respect to the zero axis, implying an anharmonicity in the related trapping potential. Indeed the harmonic oscillator with linear damping predicts a symmetric envelope and a constant angular frequency of the oscillation upon time. In our experiments, the potential anharmonicity is also witnessed by the time-evolution of the angular frequency $\omega(t)$, which is reported in Fig.\ref{fig4}c for the damped motion of Fig.\ref{fig4}b. Each value of $\omega(t)$ is obtained by analyzing the oscillation motion over 10 oscillations. The measured angular frequency is not constant but follows an exponential time evolution. We analyze this behavior by adopting a simple Duffing model with damping:

\begin{equation}
\ddot{\Theta}(t)+\lambda\dot{\Theta}(t)+{\omega_{{\rm 0}}}^2{\Theta}^2(t)+B{\omega_{{\rm 0}}}^2{\Theta}^3(t)= 0 
\end{equation}

with B$<$0 the Duffing coefficient. To deal with this non-linear equation, we propose a mathematical Ansatz inspired by our experimental results. We inject the following expression for the phase evolution 

\begin{equation}
\Theta(t)= A \exp({-\frac{\lambda t}{2}})\cos({\omega(t) t})
\end{equation}

in the equation and try to solve for $\omega(t)$. To that aim, we make several simplifications which are again validated by our experimental results. These simplifications, valid for any time t of our experimental analysis, are the followings

\begin{equation}
\lambda<<\omega(t),
\dot{\omega}(t) t << \omega(t),
\ddot{\omega}(t) t << \dot{\omega}(t)
\end{equation}

from which we obtain a simplified equation for the time evolution of $\omega(t)$, once the terms in $\lambda\omega$ are disregarded with respect to the terms in $\omega^2$

\begin{equation}
\frac{\omega (t)^2}{\omega_{{\rm 0}}^2}= 1+BA^2\exp({-\lambda t})\cos^2({\omega(t) t})
\end{equation}

At this stage, we now integrate over a period of the cosine, considering that $\omega(t)$ does not evolve at this time scale. This last step allows to smooth out the rapid time evolution and obtain the correct slow evolution of $\omega(t)$ in the form of

\begin{equation}
\omega (t)^2=\omega_{{\rm 0}}^2 (1+\frac{BA^2}{2} \exp({-\lambda t}))
\end{equation}

This is the form employed now to obtain the fit in Fig.\ref{fig4}c, with $\lambda$ extracted from the envelope evolution measured in Fig.\ref{fig4}b, and B taken as an adjustable parameter. The agreement with experimental data is very satisfactory considering the simplified mathematical solving of our non-linear model. This agreement further confirms the nonlinear dynamics of magnets levitating above superconductors, an aspect that would need to be considered for high precision experiments.\\

\par

In this work, we have focused on the rotation motion of millimetre-sized magnetic spheres. The studied motion have a very large amplitude. During hour-long rotation levitating above a superconductor,  a point on the sphere surface would typically be displaced over at least several tens of meters of curve coordinate. On the other hand, the smallest amplitude of motion that can be detected in the present experiments is commensurate with the mark spot size, on the order of the  millimeter. This size scale is still close to the magnets dimensions, a scale at which the mechanical motion is not expected to be governed by linear couplings. This is put under light in our experiments where a non-linear dynamics is revealed in several different aspects of the rotation motion. We also measured the mechanical damping associated to the dissipative pinning of vortices in the superconductor, which also lends itself to non-linearities \cite{Moon94,Matsushita07}. For all these reasons, there is a clear need for measuring the rotation of levitating sphere magnets in a purely linear regime, where best performances in terms of precision oscillator are expected. This could potentially be reached in a fluctuation-dissipation regime where the mechanical energy would only stem from dissipation, but this regime is currently not accessible to our observation. This calls for another level of sensitivity in our measurements, to resolve directly the Brownian mechanical motion of these levitating systems. Optomechanical cavity detection techniques are a natural candidate for reaching this regime, and will open a route for the optomechanics of macroscopic levitating objects.

\section*{Acknowledgments}
We thank Julien Gabelli for discussions on the physics of magnetic levitation.

\end{document}